\documentclass{ws-p10x7}

\def\PLB{{\em Phys. Lett.}  B}

\def\PRD{{\em Phys. Rev.} D}

\newcommand{\lsim}{\raisebox{-0.5mm}{$\stackrel{<}{\scriptstyle{\sim}}$}}

\begin{document}

\title{Deeply Virtual Compton Scattering at HERA}

\author{Laurent Favart}

\address{ {\rm On behalf of the H1 and ZEUS Collaborations} \\
Universit\'e Libre de Bruxelles, I.I.H.E., Belgium\\
E-mail: lfavart@ulb.ac.be}

\twocolumn 
\maketitle

\abstract{Results on Deeply Virtual Compton Scattering
at HERA measured by the H1 and ZEUS Collaborations are presented. 
The cross section, measured for the first time, is reported for
$Q^2 > 2\,{\rm GeV}^2$.}

\section{Introduction}
\begin{figure}[htbp]
 \begin{center}
  \epsfig{figure=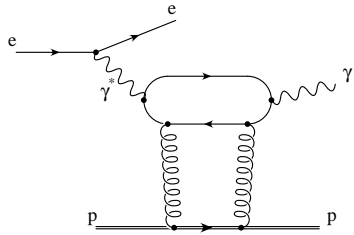,height=3.2cm}
  \caption{DVCS process.}
  \label{fig:diag1}
 \end{center}
\end{figure}

Along these last years, the exclusive vector meson production (as $\rho$
and $J/\Psi$)
has been studied extensively at HERA and has provided very
interesting results, in particular, testing for the domain of
applicability and the relevance of perturbative QCD in the field of
diffraction (see e.g.\cite{diffVM,PM}).
Here, we report the first analyses of a similar process, the  
Deeply Virtual Compton Scattering, consisting in  
the hard diffractive scattering of a virtual photon off a proton 
(Fig.~\ref{fig:diag1}).
The \mbox{DVCS} process offers a new and 
comparatively clean way to study diffraction at HERA. In comparison
to vector meson production it avoids large uncertainties 
on the theoretical predictions due to the 
meson wave-function. The largest interest comes from the 
access it gives to the skewed parton distributions of the proton\cite{rad}. 
\\

\begin{figure}[tbp]
 \vspace*{1.5cm}
 \begin{center}
  \epsfig{figure=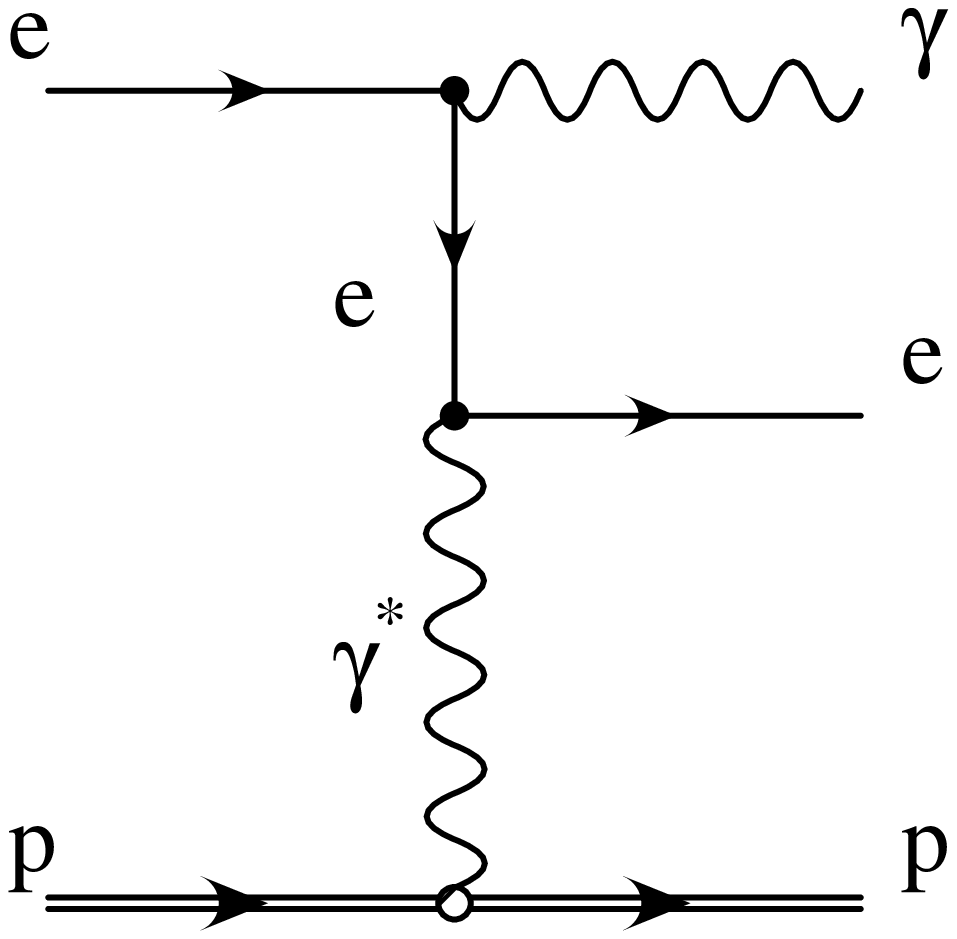,height=2.6cm}
  \hspace*{0.5cm}
  \epsfig{figure=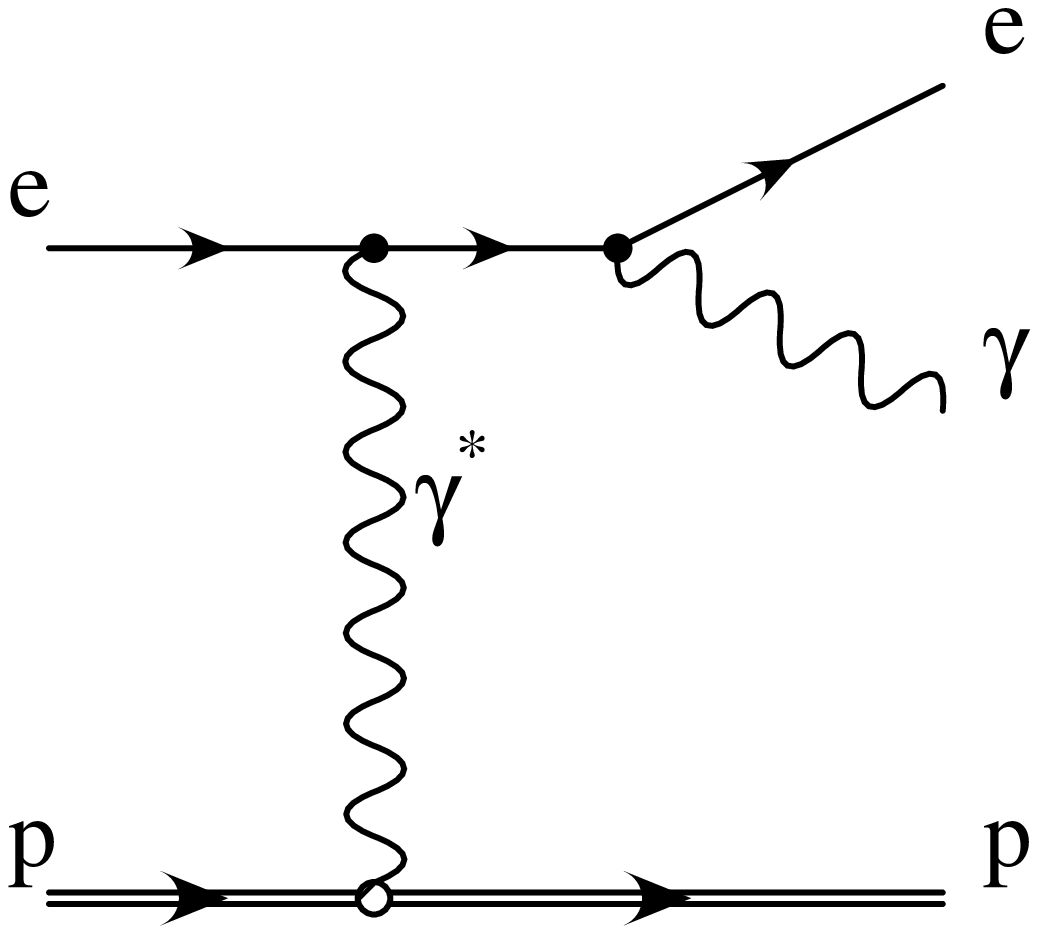,height=2.6cm}
  \caption{Bethe--Heitler process.}
  \label{fig:diag2}
 \end{center}
\end{figure}

The DVCS process contributes to the reaction 
$e^+ p \rightarrow e^+ \gamma p$, whose total cross section is 
dominated by the purely electromagnetic Bethe--Heitler process 
(Fig.~\ref{fig:diag2}).
\\

Since the virtual photon is scattered 
onto mass shell in the final state it is necessary 
to transfer longitudinal momentum from the photon to the proton,
i.e. forcing a nonforward kinematic situation\cite{rad,dvcskin}. 
In the 
picture of the two gluon exchange the fractional momentum of
the gluons are therefore not equal, which implies the 
involvement of the skewed (or nonforward) parton distribution. 
\\

In presence of a hard scale the DVCS process can be completely 
calculated in perturbative QCD. 
In the present case, the photon virtuality $Q^2$, above a few ${\rm
GeV}^2$, insures the presence of a hard scale. 
LO QCD calculations exist, based on the two gluon exchange 
model\cite{ffs}.
The factorization theorem in perturbative QCD having been 
demonstrated\cite{dvcskin,fact},
the DVCS process provides a unique way to extract these 
distributions from experimental data through the measurement 
of the asymmetry of the photon azimuthal angle distribution due to the
interference with the Bethe--Heitler process\cite{fre}. 
\\

 From an experimental point of view, HERA kinematic enables the
DVCS process to be studied 
in a large range in $Q^2$, $W$ and $t$ and offers the
possibility to study this diffractive mechanism in detail. 

\section{Analysis strategy}

Around the interaction region both experiments, H1 and ZEUS, 
are equipped with 
tracking devices which are surrounded by calorimeters.
Since the proton escapes the main detector through the beam pipe only 
the scattered electron and photon are measured. Therefore the
event selection is based on demanding two electromagnetic clusters,
one in the backward and one in the central or forward part of the 
detector ($\theta \lsim 140^o$ - the backward direction 
($\theta=0$) is defined as the direction of the 
incoming electron). If a track can be reconstructed it has to be
associated 
to one of the clusters and determines the electron candidate.
To enhance the DVCS contribution in comparison 
to the Bethe--Heitler process the phase space has to be restricted
by demanding the photon candidate in the
forward part of the detector.
\\

The H1 analysis selects more specifically the elastic component by using, 
in addition, detectors which are placed close to the 
beam pipe and which are used to identify 
particles originating from proton dissociation processes. 

\section{Results}

\subsection{ZEUS}
\begin{figure}[ht]
 \epsfxsize120pt
 \begin{center}
  \epsfig{figure=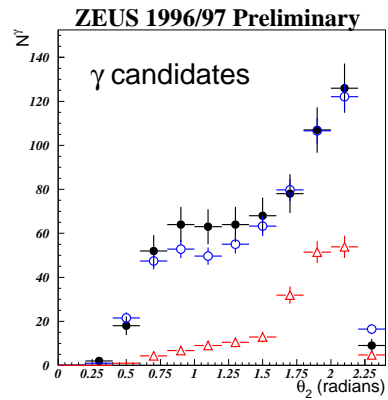,height=5.2cm}
  \caption{Distribution (uncorrected) of the polar angle of the
         photon candidate with an energy above $2\,{\rm GeV}$.
         Data correspond to the full circles. The
         prediction for the Bethe--Heitler process is indicated by the
         open triangles. The prediction of Frankfurt et al. based on
         calculations including the Bethe--Heitler and the DVCS process
         is shown by the open circles.}
  \label{fig:zeus}
 \end{center}
\end{figure}

The first observation of the DVCS process was reported by the ZEUS
collaboration in 1999\cite{saull}.  In the analysis a photon virtuality
$Q^2 > 6\,{\rm GeV}^2$ is demanded. In Fig.~\ref{fig:zeus} the polar
angular
distribution of the photon candidates is shown. A clear signal
above the expectations for the Bethe--Heitler process is observed.
The LO calculation including the DVCS and the Bethe--Heitler processes 
achives a good description of the experimental data.
A clear DVCS signal is still seen after the photon energy cut is
increased.
Also a shower shape analysis of the calorimetric clusters 
was performed that shows that the signal
originates from photons and not from
$\pi^0$ background.

\subsection{H1}

In the H1 analysis, the DVCS cross section is measured in the
kinematic region:
$2 < Q^2 < 20\,{\rm GeV}^2 $,
$ |t| < 1\,{\rm GeV}^2$
and
$30 < W < 120\,{\rm GeV}$.
The proton
dissociation background has been estimated at around 10\% and subtracted 
statistically assuming the same $W$ and $Q^2$ dependence as for the elastic
component.
The acceptance, initial state radiation of real photons and
detector effects have been
estimated by MC to extract the elastic cross section.
\\

\begin{figure}[htbp]
 \begin{center}
  \epsfig{figure=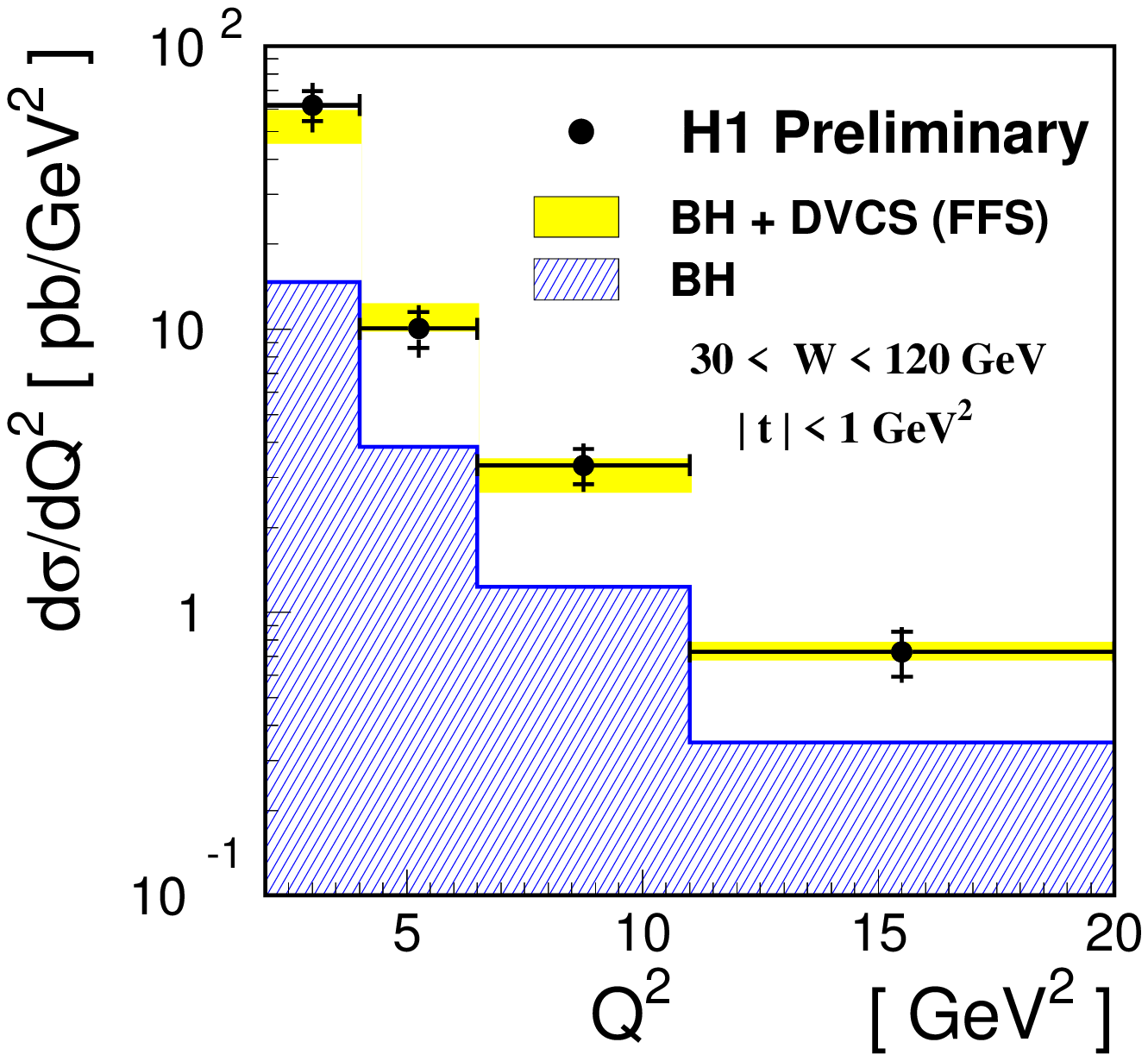,height=6.0cm}
  \epsfig{figure=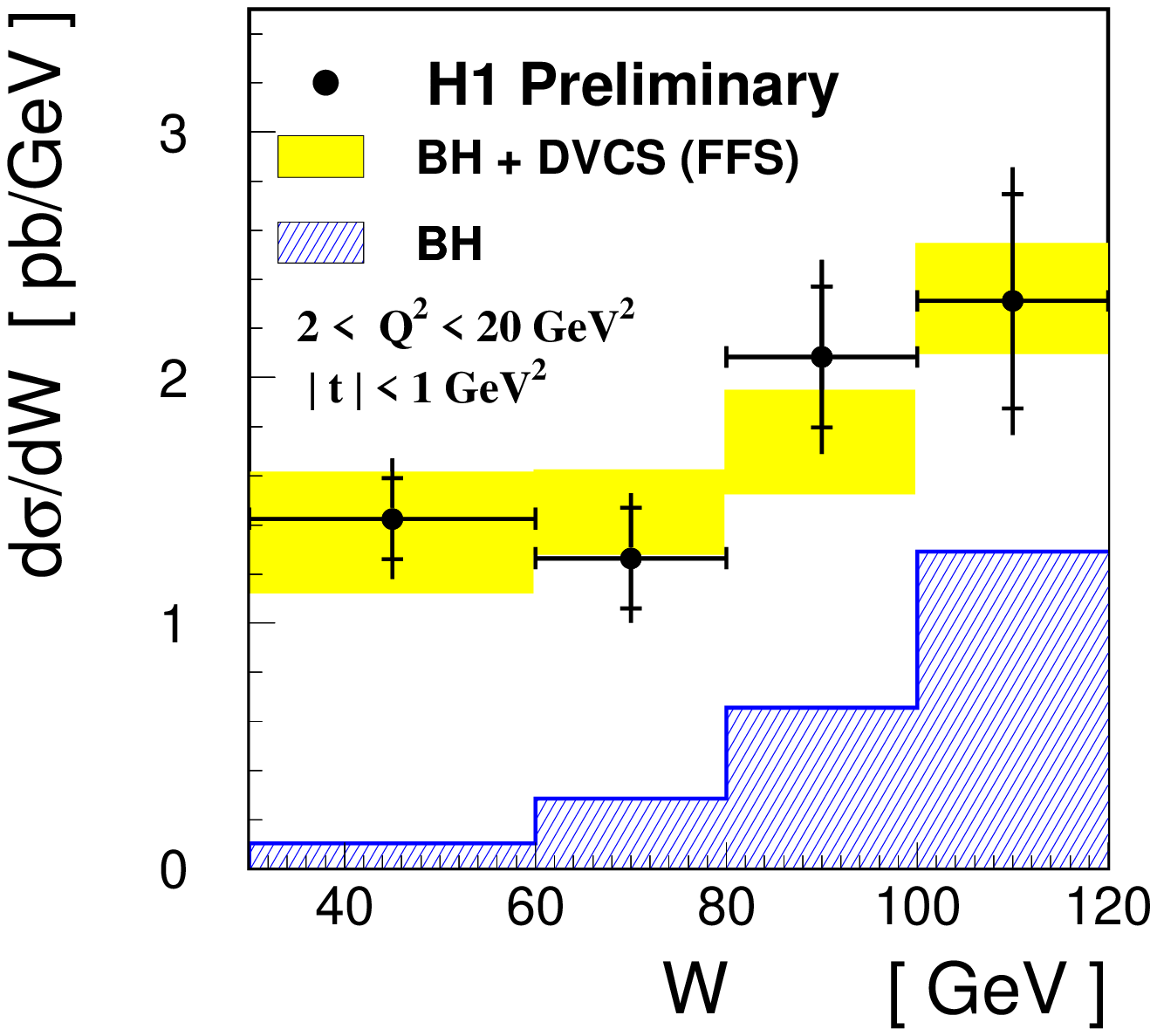,height=6.0cm}
  \caption{The measured cross sections of the reaction
  $e^+ p \rightarrow e^+ \gamma p$ as a function of
  $Q^2$ and $W$ are shown and compared to theoretical predictions.
  The uncertainty in the theoretical prediction, shown here as a shaded
  band is dominated
  by the unknown slope of the t-dependence of the DVCS part of
  the cross section, assuming $7 < b < 10\, {\rm GeV}^{-2}$}
  \label{fig:h1}
 \end{center}
\end{figure}

In Fig.~\ref{fig:h1} the differential cross sections
as a function of $Q^2$ and of $W$ are shown. The data
are compared with the Bethe--Heitler prediction alone and with
the full calculation including Bethe--Heitler and DVCS.
The description of the data by
the calculations is good,
in shape and in absolute normalization when a $t$ slope is chosen between
7 and 10~GeV$^{-2}$.
\\

It is important to notice that, at the LO, the interference term cancels when
integrating over the azimuthal angle of the final state photon (as in the 
differential cross sections in $Q^2$ and of $W$).

\section{Conclusion}

The DVCS process has been observed by the H1 and ZEUS Collaborations.
Cross section measurements have been presented for the first time.
The experimental results are well described by the
calculations of Frankfurt et al.

\end{document}